# Data mining and Privacy in Public Sector using Intelligent Agents

Authors:

Max Voskob (max.voskob@paradise.net.nz)

Contributors:

Rob Howey (rhowey@msi.net.nz)

Nick Panin (Nick.Panin@mcs.vuw.ac.nz)

Release:

26-Nov-03

## Abstract

The public sector comprises government agencies, ministries, education institutions, health providers and other types of government, commercial and not-for-profit organisations. Unlike commercial enterprises, this environment is highly heterogeneous in all aspects. This forms a complex network which is not always optimised. A lack of optimisation and communication hinders information sharing between the network nodes limiting the flow of information. Another limiting aspect is privacy of personal information and security of operations of some nodes or segments of the network. Attempts to reorganise the network or improve communications to make more information available for sharing and analysis may be hindered or completely halted by public concerns over privacy, political agendas, social and technological barriers.

This paper discusses a technical solution for information sharing while addressing the privacy concerns with no need for reorganisation of the existing public sector infrastructure. The solution is based on imposing an additional layer of Intelligent Software Agents and Knowledge Bases for data mining and analysis.

This paper uses New Zealand public sector environment as case study examples.

# Table of Contents

# Introduction

## Background information

The public sector has a mobile and complex network structure that may be hard to pin-point with a satisfactory degree of accuracy. The communications between the nodes of the network have several major information flow avenues complemented by sporadic ad-hoc connections. E.g. NZ Inland Revenue Department and NZ Statistics Department collect significantly more information from other nodes [IRD 1] [STATS 1].

From an information flow point of view, these nodes are closed systems with distinct system boundaries. The nodes have their own privacy rules, traditions and practices that form the foundation of the trust between the citizens and the node. Citizens use public services and entrust some personal and often sensitive information to the service provider. Citizens do not expect this information to be disclosed or shared with anyone outside of the public sector or even any other node except in the rare cases of public safety or when law enforcement may get involved.

This picture is characteristic to many modern democracies and greatly contributes to the democracy itself. [OECD 1]

On the other hand, information availability is essential to decision making and development of the society. This dilemma has not yet been resolved to the satisfaction of all the players.

There is always an option to enforce information sharing between the nodes of the public sector network through policies and administration. Even if the technological issues of increasing information availability can be resolved there is always some opposition from privacy activists and the general public. [TIA]

This paper discusses a solution for "information availability vs. privacy" dilemma with help of computer technology to the satisfaction of all the players. However, the scope of this paper is limited to the initial discussion of general ideas and most obvious issues arising from them.

## Traditional information sharing and analysis in public sector

Traditionally, information sharing is a top-down process with nodes guarding their information and releasing it to anyone outside the node only when it is required under the law. This itself presents a problem to the involved players (see Figure 1).

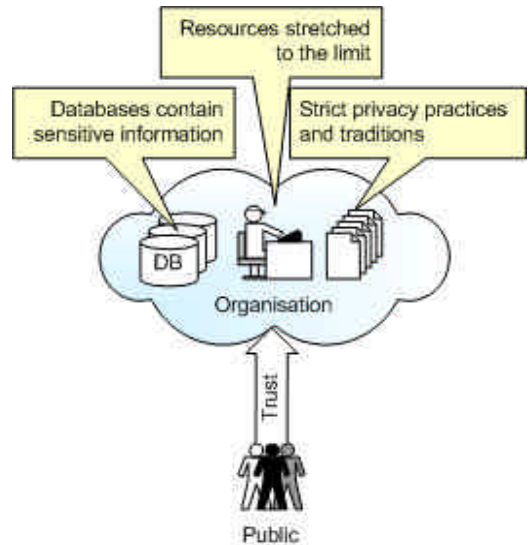

**Figure 1: Node structure**

The problems lie at both - operational and policy - levels. Privacy issues arise mainly at the policy level when the operational level have to find the resources to provide the information and solve the technological issues to make the information available.

The simplest way to make the information available is to open up database access for other nodes as illustrated in Figure 2.

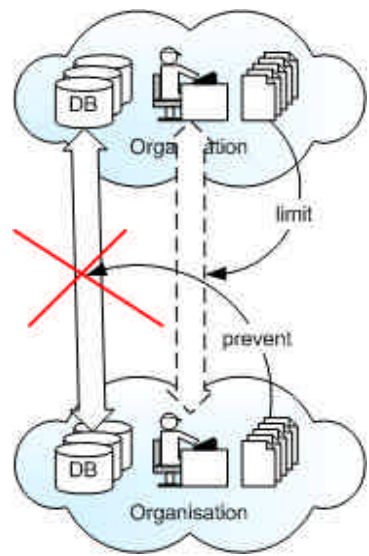

**Figure 2: Information sharing options**

A direct access from one node to the data of another node is not always possible due to privacy policies and practices that may prevent such access for anyone from outside the node. Another common channel for information sharing is via humans - members of the nodes. E.g. a member of one node sends a request for information to another node and receives back a report prepared by a member of that node with some information filtered out due to security or privacy reasons.

## Cluster vs. Individual

Privacy policies of a democratic society guard the interests of an individual, identifiable group of individuals (e.g. a club), business or an identified group of businesses (e.g. a lobbying organisation). For the purpose of this document they all will be called "entities".

According to multiple researches, citizens generally oppose any intrusion into their private information or details without their consent. At the same time, entities do not oppose when the data they provided to the public sector is being used for some analytical purposes without identifying particular individual entities. E.g. "when the statistical result separates the information from the person" [Trivellato].

Data analysis of sensitive private information that identifies clusters, not individual entities does not normally raise privacy issues.

## Case studies

### Tax evasion

It is a criminal offence to evade taxes and most developed countries have powerful IT systems in place to check up on tax payers and reconcile their expenditures, earnings and savings. Information sharing in this area is enforced by law and may be considered of being at an advanced stage compared to other areas.

### Child care and abuse

The number of players involved in this area of interest may be very large. When a strong indication about children maltreatment and abuse comes from one particular source it is often too late – the abuse has already taken place. It is one of the state's responsibilities to prevent the abuse before it happens. This can be done through information gathering and timely recognition of patterns.

For example, information of some interest may come from the Police, school, Child Youth and Family, Work and Income, health providers, health insurers, victim support organisations, charities, counselling organisations, etc. Some of these players are government agencies, some are contractors to the government, some are commercial organisations and some are non-commercial and non-government.

What kind of information can it be? Multiple researches into patterns of child abuse and maltreatment have been carried out, but the patterns evolve along with changes in our life and environment. It is hard to predict all possible patterns in advance.

In the ideal world, all the information from the players listed above can be made available to some kind of analysis to determine the patterns, recognise them and react to particular cases as soon as a pattern emerges. For example frequent visits by the police to domestic violence incidents, low performance at school, missing classes, frequent visits to different doctors with minor injuries, phone calls to victim support or youth suicide prevention lines may all be enough for early intervention by the state.

## Ad-hoc solutions

Ad-hoc solutions are common, but often they present only a half-measure and have a patchy nature. Usually changes are brought by policy makers in response to high media profile cases. An example of an ad-hoc solution can be an attempt to introduce a compulsory reporting of all visits to doctors made by children with signs of possible abuse (cuts, bruises, broken bones, internal injuries) regardless of its severity. [OUT].

This type of reporting does help identify child abuse cases, but it involves only one type of player – health providers. There is also no guarantee that all cases are being reported.

## Constraining factors

There are a number of constraining factors that limit information sharing between the nodes of the public sector network. Privacy is one of them as was discussed in the previous section. The other factors are:

- Limited resources – a node has not resources available to spend any time on preparing or gathering information for other nodes

- Node's terms of reference – a node may have specific terms of reference that allow the node to ignore some matters that may be of a serious concern to other nodes. E.g. a compliant to the Police about an abusive behaviour of a teacher may involve no criminal consequences, but it may be of a great interest to a professional disciplinary body.

- Personal agendas and beliefs of individuals – a doctor may believe that smacking children is not that bad and does not report minor cases to the authorities.

Changing the environment to remove all constraining factors does not seem practical. From a technical perspective the problem can be split in two:

- Information push – the pushing node must have a stimulus to do so whether the stimulus comes as an obligation or a reward.

- Information pull – the pulling node must be able to access the information in question in a manner that satisfies the purpose of obtaining such information.

Information push is constrained by the factors discussed previously in this section.

Information pull is constrained by technical and privacy issues. A possible solution to overcome these issues is discussed in the following sections of this paper.

## Ideal solution

An ideal solution to the information sharing problem would be a creation of one virtual information space where all the data can be mined and analysed without raising privacy issues from citizens.

This ideal solution would empower the society with greater understanding of the self helping identify problems, issues, failures and successes of many aspects of our socio-economical lives.

As it is a dream of every executive decision maker to have "information at their fingertips" it is also a dream of every policy maker. Unlike business executives, policy makers cannot integrate or consolidate all the public sector data in one database for mining and analysis due to the numerous reasons discussed previously in this paper. However, the following sections of this paper discuss a possible technical solution in detail to give the policy makers just that kind of information availability.

## Business requirements

The general requirement is to maximise information availability for analysis by third parties from outside the node without disclosing details of any individual entity to the analyst.

- Node's data about entities should be available to an undefined number of trusted third parties for analysis.

- Information about an individual entity should not be disclosed to any third party.

- Analysts should be able to reconcile information about individual entities coming from different nodes without being able to identify those entities individually.

# Technical solution

## General outline

The technical solution discussed in this paper is based on the use of Intelligent Agents, Semantics, Data Mining and Data Analysis. The solution comprises intra-node and extra-node layers.

At the intra-node layer, an Intelligent Agent gathers relevant data from the node's databases, filters it and puts it in an separate data storage for further analysis. The activities of the Agent are governed by some external and evolving rules. These activities are constrained at the data access level by the node's administration to preserve the node's sensitive data from being accessed.

At the extra-node layer, an Intelligent Agent accesses the data storages created by intra-node Agents and performs necessary data analysis. The output results of the data analysis come out in form of reports or alerts highlighting patterns, issues or risks.

There is no human involvement at either level of data access and analysis, except for configuring the agents. Humans cannot see the data that is being analysed - they may only see results of the analysis that contain no details identifying individual entities.

The Agents rely on Knowledge Bases and Rules Databases created by humans to mine, map, reconcile and analyse the data.

## Issues, risks and concerns

The following is a list of issues, risks and concerns with possible mitigation actions:

1. Why would nodes allow deployment of alien software into their environments?
- Social responsibility
- Contract obligations
- Required by law
2. Reliability of the Agents
- Deployment of open source agents
- Rigorous testing and demonstration of the reliability
3. Impact on the node's IT resources due to an increased workload (hardware, IT personnel)

- Proper design of the agent architecture for minimal intervention and maximum autonomous functioning
- Remote administration and support by a dedicated team
- Low prioritisation of agent's access to the node's databases (tasks can be done only if they do not impact on the normal node activities)
4. Threat of malicious activities by the Agents
- Deployment of open source agents
- Use of Public Key Infrastructure [PKI].
5. Security risks from providing external interfaces
- Use of traditional defence and prevention mechanisms such as firewalls, etc.
6. Difficulties with reconciling data from multiple data sources from the same or multiple nodes
- Use of semantics for data and information mapping
- Use of common vocabularies where possible
- Use of shared data matching and mapping services, e.g. name and address recognition and matching
7. Individual entities and related to them events may be identified by people from outside of the node hosting the data
- Deployment of open source agents that provide only generalised reports and alerts
- Extra-node agents do not store processing data in persistent form so that it cannot be recovered later
8. Malicious agents may be deployed or replace authorised agents to harvest or analyse data
- Use of Public Key Infrastructure [PKI].
9. Rules and knowledge bases may be modified or corrupted for malicious purposes
- Use of Public Key Infrastructure [PKI].
- Access control
- availability for the rules and knowledge bases in read-only format for audit

# Technical solution details

## Intra-Node layer

The intra-node part of the solution is illustrated in Figure 3. The cloud in the picture represents a node with some external interfaces.

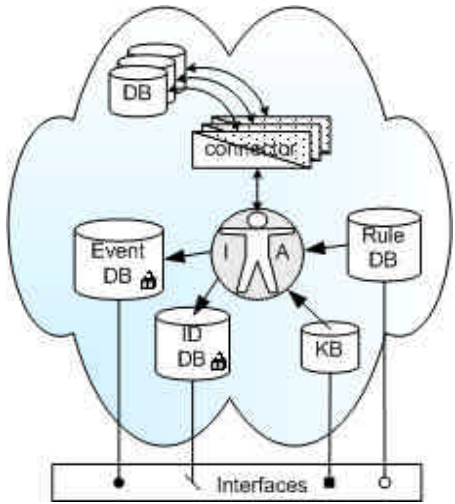

**Figure 3: Intra-node Intelligent Agent**

**IA**

Intelligent Agent
A piece of software that mines node's databases and extracts data about entities and events as specified in the rules

**DB**

Node's databases, content management systems and other data storages with relevant information about entities and events

**Connector**

A piece of software that allows the IA to connect to different types of DBs. Connectors may provide additional security and filtering features to limit access to sensitive data.

**Rule DB**

Rules Database
Contains data mining rules specific to the node's environment

**KB**

Knowledge base
Contains formal knowledge representation in a form of ontology for configuration, data mining and mapping

**Event DB**

The output database with data about events in relation to individual entities
Encrypted

**ID DB**

An identity database for individual entities
Used to match the identities of individual entities from different nodes without revealing identity details
Encrypted

**Interfaces**

Web service interfaces exposed to the outside world. Some security and access control measures have to be applied.

The process of deploying, configuring and using the IA is described in the following subsections.

### Deployment

The IA is an application that can be deployed to any environment and act autonomously. The IA connects to the node's databases through special connectors that provide an abstraction level for data access.

The IA is deployed with the Rule DB, KB, ID DB and Event DB as a software package.

The node administrators set up access to the node's databases and provide other security permissions, e.g. a connection to the local proxy-server to use and provide web-services.

### Configuration

Rule DB contains a set of rules in a form of ontology or another formal language. It describes what type of information the agent is required to collect. E.g. an agent deployed to the Police environment may collect information on calls to domestic violence incidents only. The Rule DB would have a set of rules that explain it to the agent.

The Knowledge Base (KB) provides semantics for the information related to the event that is being logged – domestic violence in this case. Also, the KB must provide prior knowledge of some common terms, e.g. describe notions of person, name, address, event, incident, police, phone number, etc. and relationships between them.

The newly deployed IA has no prior knowledge of the data models used in the node's databases. There are two options to discover the semantics of the data: analyse the data or use semantics provided by administrators. Extensive data analysis may pose some security risks or may be still incorrect so we assume that semantics are provided by the administrators in a form of mapping between database fields and the Knowledge Base. E.g. a DB admin may be presented with a GUI screen that has the data model of the database and ontology from the KB. Dragging and dropping fields to the ontology nodes will build a semantic map for the agent. Given the map, the agent knows what fields are of interest, their semantics and how they relate to each other. [ERCIM]

When the mapping is complete, the agent may begin to collect information.

### Information gathering

The agent gathers data from the node's databases and puts it in the Event Database. The Event DB may have a generic data model with semantic descriptions attached to it so that whoever mines the Event DB may understand the meaning of the data without any prior knowledge.

All identity information (e.g. names, address, phone numbers, account numbers, etc.) is stored in the Identity DB for security reasons. The identity information can be still reconciled with the event information at any time.

### Example

The Police DB contains information about a number of phone calls from Mr XYZ living in 32 ABC St that there were heated arguments in the neighbouring flat and the children were screaming. The Police attended the dispute, but no charges were laid. The agent logs this information in the Event DB as a list of records in form of "who-when-did-what" and the identities of all involved are stored in the identity database.

### Interfaces and information provisioning

The Event DB, Rule DB, Identity DB and Knowledge Base expose some interfaces to provide data to consumers from outside the node.

The Event DB and Identity DB contain sensitive information that should be encrypted. The data from these storages can be provided to the extra-node Intelligent Agents only.

The Rule DB and Knowledge Base can be queried by authorised external players to determine the rules for the agent and gather the knowledge available to the agent. These data storages may be also encrypted or the access can be restricted in some other ways.

## Extra-node layer

The extra-node layer has a similar configuration to the intra-node and relies on the same players: IA, Rule DB, KB, Event and Identity DBs. The extra-node agents reside outside node boundaries as illustrated in Figure 4.

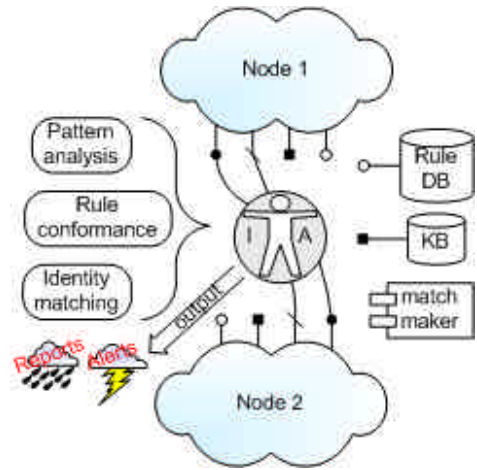

**Figure 4: Extra-node Intelligent Agent**

Extra-node agents connect to multiple Event and Identity DBs from different nodes to harvest the event and identity data. One of the most important tasks is data matching – events related to one entity from one Event DB have to be matched with events related to the same entity from other Event DBs.

When this data matching/reconciliation process is completed the agent may analyse the resulting dataset and provide a report or alert based on the findings. The following subsections describe this process in detail.

### Deployment

An authorised player interested in analysing some aspects of the public sector may deploy an extra-node agent to perform a specific analysis task. The agent may be tailored to perform specific analysis tasks, but it is expected to be a mainly generic piece of software.

### Configuration

The agent uses a shared Rule DB and Knowledge Base for its configuration and as a source of knowledge. The agent has to be provided with connection details to access the intra-node Event DBs and KBs. The connection details may include certificates and other security and access control details.

## Information gathering and matching

The agent connects to the specified Event and Identity DBs. The first task would be data matching and reconciliation. E.g. Event DB A contains data about phone calls made to Child Youth and Family help line including name and addresses of the children in question. On the other end can be a set of school records of missed classes with slightly different identity details, but about the same individuals. Direct one-to-one matching is unlikely to paper – some more elaborate procedures are needed in this case.

To assist the matching process, all the nodes have to use the same shared semantics so that the meaning of data from node A can be matched to meaning of data from node B.

The data matching service can be shared by all extra-node agents as a service.

When matching is complete and the data is reconciled the agent has a complete data set in hands to analyse.

## Data analysis

Extra-node agents may perform all sorts of data analysis, but some of the most obvious are:

- Pattern identification, when the agent identifies patterns in the system and entity behaviour.

- Rule conformance, when a citizen's actions go against some rules or law, but it cannot be easily identified in other ways. E.g. a rule may say that the principal occupant of a state provided flat must permanently reside at that address. If a principal occupant of such a flat in Wellington makes regular visits to a doctor in Auckland then the rule is likely to be broken.

- Ad-hoc analysis, when someone wants to understand trends or dependencies or perform a one-off analysis task.

Results of the analysis may come in form of reports or alerts or just be dumped to a database for further consideration.

## Example

The Ministry of Social Development (MSD) may be interested in following on the lives of migrants after arrival to New Zealand. This may include numerous aspects of every day life and unidentified list of service providers. Assume that most of the public sector service providers have intra-node agents and their data is available for harvesting and analysis.

To perform this analysis the MSD may deploy an extra-node agent that takes the data from the New Zealand Immigration Service first and then crawls around the public sector network to visit all service providers for information. Given the identities of the migrants it will be a feasible task to find out about many aspects of their lives, such as:

- What education they have

- What professions they have

- Who do they work for

- Where they live

- How often they move

- Do they buy a house or rent

- What schools their children go to

- How often they go to their home countries

- How often they go to see a doctor

- What medical problems they have

- What their income is and how it changes over time

- etc.

Analysis of this kind may give an unprecedented understanding of the adaptability of different categories of migrants, their social and economic impact. It may lead to a smarter immigration and settlement policies.

# The big picture

## Multiple agents environment

In An ideal environment multiple agents may be freely deployed by any authorised player as illustrated in Figure 5. An abstract analogy of the environment can be a beehive where the system works, but you may let your own bee into the system to perform your specific analysis tasks.

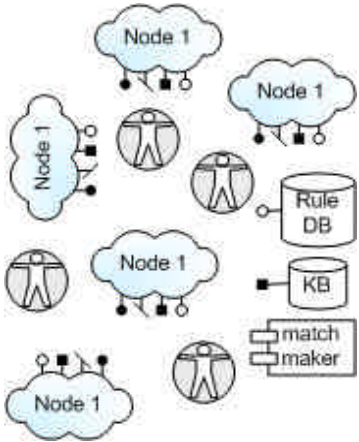

**Figure 5: Multi-agent environment**

It is unlikely that the system can be completely open to any player and the extra-node agents may physically reside anywhere. Due to security vulnerability of an open environment the extra-node agents may be required to reside in a secure dedicated environment that would guarantee that only the authorised players deploy agents, the agents do not perform malicious actions, the information does not leak outside of the secure environment, etc.

## Privacy issues

The use of Intelligent Software Agents eliminates the weakest link in any secure system – humans. Once the agent is deployed it is completely autonomous. It accesses the data, analyses it and returns a report, alert or any other data that has no direct link to an individual. Humans may use the information they get from the agents to retrace the individuals, but they have to do it using conventional channels of inter-agency communications. E.g. they may to raise an issue with policy makers to request that some Police data has to be matched with data from two other service providers in real time for a timely response to the most common child abuse patterns with identification of the involved individuals.

### Privacy requirements

This solution has to meet strict privacy requirements to win public confidence. Some of the requirements are:

- The agents are developed as open source solutions with all the source code available

- The design of the agents prevents any human intervention except for provisioning of rules and knowledge

- All the data provided by the intra-node agents is secured against unauthorised access

- The extra-node agents do not store the data harvested from the nodes the way that humans may take hold of it

- The design of the extra-node agents prevents any disclosure of information directly identifying individual entities

- The rules and knowledge provided for the agents are available for public or other audit

- IT technologies are used wisely in an open manner without compromising the overall security and reliability of the solution

This list is by no means comprehensive at this stage. As the solution matures and IT technologies evolve it will obviously evolve further.

The openness of the solution is the key factor to winning public confidence that it is not about a Big Brother spying on all of us – it is about helping us gain knowledge and understanding of our socio and economic lives.

## Dependant technologies

This solution depends on some emerging technologies such as web semantics, ontologies, automatic data matching and reconciliation, intelligent agents and some others.

The major dependencies are The Semantic Web, Web Service Architecture and Intelligent Software Agents. These are generalised streams of technologies that comprise a number of specific technologies, approaches and visions. Their levels of maturity vary from a research stage to available implementations by mainstream software vendors.

## Feasibility

The solution proposed in this paper is not available as an out of the box product at present. Probably it will not be available for some years to come.

In fact, the feasibility of this solution depends on a number of key factors:

- availability of the dependant technologies
- demand from the public sector for such solutions to create a market for commercial vendors to move in
- wiliness of the policy makers to endorse this kind of solution
- investments into more academic research of the aspects of the solution and related technologies

This list is preliminary as this paper presents only an initial discussion into the matter.

## Further discussion

The authors of this document invite interested organisations and individuals to consider this solution and take part in the discussion.

## About the authors

Max Voskob is an XML and Interoperability consultant with MSI Business Solutions NZ Ltd., Wellington, NZ.

Rob Howey is a Data Quality and Analysis consultant with MSI Business Solutions NZ Ltd., Wellington, NZ.

Nick Punin is a Computer Science student with Victoria University, Wellington, NZ

# References and related work

**Topic Maps**

ISO/IEC 13250:2000 "Topic Maps" standard [ISO13250]

http://www.topicmaps.org/

**OWL**

Web Ontology Language by W3C Web-Ontology (WebOnt) Working Group
http://www.w3.org/2001/sw/WebOnt/

**W3C Web Service Architecture**

Architecture of the World Wide Web
by W3C's Technical Architecture Group (TAG)
http://www.w3.org/TR/2003/WD-webarch-20031001/

**W3C Semantic Web**

Architecture of the World Wide Web
by W3C's Technical Architecture Group (TAG)
http://www.w3.org/2001/sw/

**PKI**

OASIS Public Key Infrastructure TC
http://www.oasis-open.org/committees/tc_home.php?wg_abbrev=pki

**IRD 1**

New Zealand Inland Revenue Department
Information Exchange
http://www.ird.govt.nz/aboutir/infoexchange.html

**STATS 1**

New Zealand Statistics Acts 1975
Exerts are available at
http://www.stats.govt.nz/domino/external/web/aboutsnz.nsf/htmldocs/Statistics+Act+1975

**OECD 1**

OECD Guidelines on the Protection of Privacy and Transborder Flows of Personal Data
ISBN 9264197192
http://www1.oecd.org/publications/e-book/9302011E.PDF

**TIA**

Total / Terrorism Information Awareness

The US Department of Defense
http://www.darpa.mil/body/tia/tia_report_page.htm

**ERCIM**

Towards Syntax-Independent B2B

ERCIM News No. 51
Special Report: Semantic Web

**QUT**

Suspected Child Abuse and Neglect Reporting Behavior
Queensland University of Technology

http://www.abusedchildtrust.com.au/content/documents/ACTPAPER-SuspectedCANreportingbehaviour.pdf

**Trivellato**

Data access versus privacy: an analytical user's perspective by Ugo Trivellato, Statistica, 60 (4), 2000
www2.stat.unibo.it/rivista/sum2000/sum4_00.htm